\documentclass[a4paper]{article}
\usepackage{color}
\usepackage{amsthm}
\usepackage{amsmath}
\usepackage{graphicx}
\usepackage{amssymb}
\usepackage{bbold}
\usepackage{esint}
\usepackage{subfigure}
\usepackage{color}
\usepackage[all]{xy}

\usepackage{anysize}
\marginsize{2cm}{2cm}{2cm}{2cm}

\def\beq{\begin{equation}}
\def\eeq{\end{equation}}
\def\bea{\begin{eqnarray}}
\def\eea{\end{eqnarray}}
\def\nn{\nonumber}

\makeatletter

\theoremstyle{plain}

  \theoremstyle{remark}

\begin{document}

\title{Toric code-like models from the parameter space of $3D$ lattice gauge theories}
\author{  Miguel Jorge Bernab\'{e} Ferreira,$^{a}$\footnote{migueljb@if.usp.br}~Pramod Padmanabhan,$^{a}$\footnote{pramod23phys@gmail.com}\\ Paulo Teotonio-Sobrinho$^{a}$\footnote{teotonio@fma.if.usp.br}  } 
\maketitle

\begin{center}
{\small $^{a}${ Departmento de F\'{i}sica Matem\'{a}tica Universidade de S\~{a}o Paulo- USP}}
\end{center}

\begin{abstract}

A state sum construction on closed manifolds \'{a} la Kuperberg can be used to construct the partition functions of $3D$ lattice gauge theories based on involutory Hopf algebras, $\mathcal{A}$, of which the group algebras, $\mathbb{C}G$, are a particular case. Transfer matrices can be obtained by carrying out this construction on a manifold with boundary. Various Hamiltonians of physical interest can be obtained from these transfer matrices by playing around with the parameters the transfer matrix is a function of. The $2D$ quantum double Hamiltonians of Kitaev can be obtained from such transfer matrices for specific values of these parameters. A initial study of such models has been carried out in \cite{p1}. In this paper we study other regions of this parameter space to obtain some new and known models. The new model comprise of Hamiltonians which ``partially'' confine the excitations of the quantum double Hamiltonians which are usually deconfined. The state sum construction allows for parameters depending on the position in obtaining the transfer matrices and thus it is natural to expect disordered Hamiltonians from them. Thus one set of known models consist of the disordered quantum double Hamiltonians. Finally we obtain quantum double Hamiltonians perturbed by magnetic fields which have been considered earlier in the literature to study the stability of topological order to perturbations.    

\end{abstract}

\section{Introduction}

{\it Topological Order} - Topologically ordered systems have gained wide attention in recent years due to some of its consequences in topological quantum computation and emergence of new phases of matter among many others~\cite{Nr}. Among the different types of systems exhibiting topological order the ones with long-ranged entangled (LRE) ground states are the ones which are thought to be most useful for quantum computation. The earliest proposals of such systems are the quantum double Hamiltonians of Kitaev~\cite{Kit1, Hag}. The {\it toric code} is the simplest example of a $2D$ lattice systems which contains anyons as low energy excitations and have degenerate LRE states as ground states, this model consists of spin variables living on the links of a square lattice. These were further generalized by the Levin-Wen models~\cite{LW} or the string-net models which described more general anyonic excitations by directly taking a unitary fusion category as inputs. These models are also quantum double models based on weak Hopf algebras as noted in~\cite{LChang} and can thus be constructed via the algorithm of Kitaev. Topological codes have also been considered on manifolds with boundary \cite{del2, del3}. Several other models inspired by the usefulness of the toric code as a stabilizer code have been constructed of which the topological color codes~\cite{TCC, TCC2, TCC3} are an example which have also been experimentally implemented~\cite{Del}.

{\it State Sum Constructions and Statistical Mechanical Models } - Both the Levin-Wen model and Toric code models can be thought of as Hamiltonian realizations of topological field theories (TQFTs) and then they can be formulated in terms of topological invariants. State sum constructions of TQFT's \cite{cfs} have been employed in realizing statistical mechanical models in the past \cite{paulo1, paulo2, paulo3}. Such methods have also been used to construct the Levin-Wen models~\cite{LW} using the Turaev-Viro invariants \cite{KuperLW} and chain-mail link invariants ~\cite{SimonLW1, SimonLW2}. Kitaev's toric code has also been related to Turaev-Viro codes~\cite{KiriToric}. The Levin-Wen model corresponds to a topological invariant called Barrett-Westbury invariant~\cite{BW} and the toric code corresponds to a special case of the Kuperberg invariant~\cite{p1}. This has been especially noted in~\cite{Kit2}. We showed this explicitly in~\cite{p1} where we embedded the $2D$ quantum double models based on an involutory Hopf algebra, $\mathcal{A}$, in an enlarged parameter space (defined later), that of the $3D$ generalized lattice gauge theories based on these algebras $\mathcal{A}$. In the special case where the algebra $\mathcal{A}$ is taken to be the group algebra $\mathbb{C}G$ of a group $G$, the generalized gauge theory can reproduce the lattice gauge theories familiar to physicists \cite{vitor}. The toric code occurs when we choose $G=Z_2$ \cite{p1}.  

{\it The Quantum Double Model} - The quantum double model of a discrete group $G$\footnote{The quantum double model is defined for more general inputs like involutory Hopf algebras and more generally weak Hopf algebras. We will only be concerned with the case of group algebras which are a particular case of involutory Hopf algebras.} is defined on a bidimensional lattice over a compact manifold $\Sigma$ of genus $g$. The degrees of freedom live on the links of the lattice and they are vectors $\vert g \rangle_l \in \mathcal{H}_l$, where $g\in G$ and $l$ represents a link of the lattice. The total Hilbert space $\mathcal{H}$ is then the tensor product of all $\mathcal{H}_l$ (for all $l$), in other words $\mathcal{H}=\mathcal{H}_1\otimes \mathcal{H}_2\otimes \cdots\otimes \mathcal{H}_N$, where $N$ is the total number of links. A basis vector $\vert \Psi \rangle$ of $\mathcal{H}$ is then of the form $\vert \Psi \rangle = \vert g_1 \rangle_1 \otimes \vert g_2 \rangle_2 \otimes \cdots \vert g_N \rangle_N$, with $g_i\in G$. The dynamics of such a model is governed by a Hamiltonian $H^{QD}$ made up of a sum of commuting operators, acting locally on the plaquettes, $p$, and vertices, $v$ of the lattice and this is given by 
\begin{equation}
\label{HDQ}
H^{QD}=-J_p \sum_p B_p - J_v \sum_v A_v\;,
\end{equation}
where $B_p$ is the plaquette operator, $A_v$ the vertex operator and $J_{p, v}$ positive numbers. These operators are both projectors and also commute with each other for all vertex and plaquettes making energy levels discrete. 


Consider $\{ \vert a \rangle:a=1,2,\cdots,k\}$ a complete set of eigenvectors of $H^{QD}$ with $H^{QD}\vert a \rangle = E_a \vert a \rangle$. If the system is put into a bath with temperature $T$ one can obtain thermodynamics properties of such a system by its partition function 
\begin{equation}
\label{HeZ}
Z^{DQ}=\sum_{a=1}^k \langle a \vert e^{-\beta H}\vert a \rangle=\hbox{tr}(e^{-\beta H})\;,
\end{equation}
with $\beta=1/k_B T$, being $k_B$ the Boltzmann constant. The matrix $e^{-\beta H}$ is called the transfer matrix. Since $H$ is given by equation \ref{HDQ}, the partition function can also be written as 
\begin{equation}Z^{DQ}(\beta, J_p,J_v)=\hbox{tr}\left(\prod_p e^{\beta J_p B_p} \prod_v e^{\beta J_v A_v}\right)\;.\end{equation}
In the special case where $J_v=0$ this partition function can be rewritten as
\begin{equation}Z^{DQ}(\beta, J_p,0)=\hbox{tr}\left(\prod_p e^{\beta J_p B_p}\right)=\sum_{\hbox{\small conf.}} \prod_p e^{\beta J_p S(p)}\;,\end{equation}
where the sum runs over all the configurations and $S(p)=+1$ if the holonomy of the plaquette $p$ is flat and $S(p)=-1$ otherwise. The function $S(p)$ is invariant under gauge transformation, and then $Z^{DQ}(\beta, J_p,0)$ is the partition function of a lattice gauge model. For $J_v\neq 0$ it can also be shown that $Z^{DQ}$ is a partition function of some lattice gauge model \cite{p1}.

{\it Partition Functions of $3D$ Lattice Gauge Theories} - We mentioned earlier that the quantum double Hamiltonian is related to the partition function of a lattice gauge model, as once we know the Hamiltonian the partition function is well defined. Now we can ask the question, if a partition function $Z$ is given, is it possible to obtain a Hamiltonian $H$ such that the equation \ref{HeZ} is satisfied? The answer for this question is no, however if $Z$ is restricted to be gauge invariant it can be done. The reason is that there is a way of building gauge invariant partition functions out of a $3D$ topological invariant called the Kuperberg invariant \cite{kuperberg}, which is based on involutory Hopf algebras. Moreover this construction allows us to obtain partition functions that are more general than the lattice gauge ones, but they are still gauge invariant. 


This partition function is parametrized by four non-physical parameters, namely $z_S,z_T\in \mathbb{C}G$ and $\xi_S,\xi_T \in \mathbb{C}G^*$, in other words $Z$ will be a function of the form $Z(z_S,z_T,\xi_S,\xi_T)$. The choice of such parameters leads to specific models. It is important to note that these parameters are not free such as for example $\beta$ in ordinary gauge theories. These parameters are fixed once the model is fixed. In particular, as it can be seen in \cite{vitor}, if the parameters are taken to be \footnote{Here $\{\phi_{+1},\phi_{-1}\}$ is the basis for the group algebra $\mathbb{C}\mathbb{Z}_2$.}
\begin{equation}z_s=z_T=\frac{1}{2}e^{-\beta}\phi_{+1}+\frac{1}{2}e^{-\beta} \phi_{-1}\;,\end{equation}
and $\xi_S=\xi_T=\epsilon$, where $\epsilon$ denotes the counit of the algebra $\mathbb{C}G$. The partition function obtained by such a choice coincides with that of a $3D$ pure lattice gauge theory with $G=\mathbb{Z}_2$. Different choices of these parameters lead to different partition functions which may not be related with gauge theories at all, however if we restrict the parameters $z_S$ and $z_T$ to be elements of the centre of $\mathbb{C}G$, whatever the partition function is, it will be gauge invariant \cite{vitor}. The action in this partition function may not be of physical interest but they will all be gauge invariant, and for that reason we will say that $Z(z_S,z_T,\xi_S,\xi_T)$ is the partition function of a {\it generalized gauge theory}. From now on we  consider $G$ to be a discrete group, and the $3D$ dimensional lattice a finite cubic lattice as a triangulation of a manifold of the form $\Sigma\times S^1$, where $\Sigma$ is a $2D$ compact manifold of genus $g$ and $S^1$ is the one dimensional sphere.

{\it Kuperberg's Construction of Transfer Matrices} - The way to build these generalized partition functions $Z(z_S,z_T,\xi_S,\xi_T)$ is by associating tensors, made up of the structure constants of the algebra, to the faces and links of the lattice leading to a very complicated tensor network that, fortunately, can be realized as the trace of a matrix $U(z_S,z_T,\xi_S,\xi_T)$ by
\begin{equation}Z(z_S,z_T,\xi_S,\xi_T)=\hbox{tr}\left(U(z_S,z_T,\xi_S,\xi_T) \right)\;,\end{equation}
where, by analogy, we can think of it as being the exponential of a Hamiltonian $H(z_S,z_T,\xi_S,\xi_T)$, i. e. 
\begin{equation}U(z_S,z_T,\xi_S,\xi_T)=e^{-H(z_S,z_T,\xi_S,\xi_T)}\;,\end{equation}
where the $\beta$ constant can be suppressed without loss of generality. The matrix $e^{-H}$ can be thought of as an operator acting on a $2D$ lattice (over the $2D$ manifold $\Sigma$), but for that we have to make distinction between the {\it timelike} and {\it spacelike} directions on the original $3D$ lattice\footnote{The terms {\it timelike} and {\it spacelike} are just used to distinguish the spacelike directions in $2D$ from the third direction which we call timelike. We still work in the Euclidean metric.}. This procedure is shown in detail in \cite{p1}, where a Hilbert space $\mathcal{H}$ is associated with the $2D$ lattice as $\mathcal{H}=\mathcal{H}_1\otimes\mathcal{H}_2\otimes\cdots\mathcal{H}_N$, with $\mathcal{H}_l\sim \mathbb{C}G$ being the local Hilbert space associated with the link $l$ of the $2D$ lattice and $N$ is the total number of links of the $2D$ lattice. The degrees of freedom are then elements of $\mathbb{C}G$ living on the links, which is equivalent to saying that the degrees of freedom are group elements, since there is a one-to-one correspondence between group elements and elements of the basis of $\mathbb{C}G$. 

{\it Models from the transfer matrix} - We then obtain the Hamiltonian by taking the logarithm of $U(z_S,z_T,\xi_S,\xi_T)$. However the Hilbert space $\mathcal{H}$ is very huge, which makes $U(z_S,z_T,\xi_S,\xi_T)$ a very huge matrix and difficult to take its logarithm. But as we have done before \cite{p1} the matrix $U(z_S,z_T,\xi_S,\xi_T)$ can be decomposed into a product of local operators acting on $\mathcal{H}$ given by
\beq U(z_S,z_T,\xi_S,\xi_T) = \prod_p B_p(z_S)\prod_l T_l(\xi_S)L_l(z_T)\prod_v A_v(\xi_T),\eeq
where $p$, $v$ and $l$ denotes plaquettes, vertices and links respectively with $T_l$ and $L_l$ operators acting on the degrees of freedom located on the links $l$ and $B_p$ and $A_v$ the plaquette and vertex operator which we previously encountered in the quantum double Hamiltonian given in Eq. \ref{HDQ}. These vertex and plaquette operators satisfy the quantum double algebra \cite{bais, SM}\footnote{The notion of quantum doubles arises in the theory of Hopf algebras where the quantum double construction is used to generate a quasitriangular Hopf algebra from a given Hopf algebra. A quasitriangular Hopf algebra is governed by a $R$ matrix which satisfies the quantum Yang-Baxter equation (QBYE). This can also be taken to be a way to generate solutions for the (QBYE) which coincide with the irreducible representations of the braid group in two dimensions. In physical terms these are anyons which are also the irreducible representations of the quantum  double algebra and hence the usefulness of the quantum double Hamiltonians in obtaining anyons in the spectrum of the theory. The reader is referred to the book on Hopf algebras where these ideas are discussed \cite{SM}.}.
Moreover the plaquette and vertex operators commute with each other for all choices of plaquettes and vertices. The link operators do not commute in general and thus switching on the parameters corresponding to them namely, $z_T, \xi_S$ will complicate the procedure of taking logarithms of the transfer matrix. Therefore we can not obtain exactly solvable Hamiltonians for an arbitrary choice of the parameters $z_S$, $z_T$, $\xi_S$ and $\xi_T$, we can only do it for those of which the local operators commute with each other. One such Hamiltonian is the quantum double Hamiltonian of Eq. \ref{HDQ} where only the parameters $z_S$ and $\xi_T$ corresponding to the plaquette and vertex operators are used. 

{\it Other models using more parameters in the transfer matrix} - Some examples include quasi-topological phases which result from a condensation of the excitations~\cite{Del2} of the quantum double phase of Kitaev. This leads to increased ground state degeneracy for the condensed phases. Examples of these phases were studied in~\cite{p1}. These phases including the quantum double phases of Kitaev were obtained when we considered the parameters $z_S$ and $\xi_T$ in the transfer matrix. In~\cite{p3} we showed that we could obtain the quantum double phases of Kitaev by  writing down models which included $z_T$ or $\xi_S$.

 {\it Identifying topologically ordered phases} - The models that exhibit topological order in $2$D can be understood by their quasi-particle content called anyons. The data that determines the phase are the ground state degeneracy, their statistics and fusion parameters \cite{Wang}. If two models have different ground state degeneracies, or different fusion rules or statistics, they are not in the same topological phase.

In this paper we take this program further by considering more parameters in the transfer matrix which were not included in~\cite{p1, p3}. We consider three types of models here. Two of them do not include $z_T$ and $\xi_S$ while the third includes them. The first two sets of models comprise of the disordered quantum double Hamiltonians of Kitaev and a new Hamiltonian which leads to ``partial'' confinement of the excitations of the quantum double phase of Kitaev. We use the term partial to emphasize the fact that the models are such that the excitations can be moved a few steps with an energy cost after which they become deconfined like in the usual quantum double models. According to the terms added we can confine the excitations for any number of steps that we wish to. We will also call these models $n$-step confined models in the text to follow. The models which include the other two parameters, $z_T$ and $\xi_S$ are the quantum double Hamiltonians perturbed by magnetic fields. These can be thought of as local perturbations to the exactly solvable Hamiltonians of Kitaev. Due to the usefulness of these models to realize fault tolerant quantum computation, it is necessary to study the stability of the topological order to local perturbations~\cite{Bravyi1, per, per2}. These models have already been considered in the literature and we write them down here just for the sake of completion and to drive home the point that they are well within the parameter space of the three dimensional lattice gauge theories. Our focus is on exactly solvable Hamiltonians like the original toric code Hamiltonian and so phase transitions are out of the scope of this paper as we will then necessarily have to move through perturbed toric code Hamiltonians which are outside the exactly solvable regime. 

The contents of the paper are organized as follows. Section 2 gives a brief review of the construction of the transfer matrix of the generalized lattice gauge theories. The section also includes an introduction to the mathematical preliminaries that go into the construction of the partition function and the transfer matrices. The algebra of operators, which include the quantum double relations between the vertex and plaquette operators, are written down. The models obtained from this transfer matrix are described in section 3. An outlook is presented in section 4.   

\section{Partition Function and Transfer Matrix of Generalized Lattice Gauge Theories}
~

The partition function of lattice gauge theory is a well known example of a classical partition function built out of local weights associated to plaquettes of an oriented 3D lattice, where the gauge degrees of freedom are elements of a gauge group $G$ living on the edges of the lattice. A configuration is a choice of an element $g\in G$ for each link of the lattice. For $G=\mathbb{Z}_2$ the gauge degrees of freedom are spin variables $\pm 1$ living on the links. The action of this model is defined by $S=\frac{1}{2}\sum_p \left(\hbox{tr}(U_p)+\hbox{tr}(U_p^{-1}) \right)$, where the sum runs over the plaquettes of the lattice and $U_p$ is the holonomy of a plaquette $p$. The partition function which describes the model is given by
\begin{equation}
\label{kaka}
Z=\sum_{\hbox{\small conf.}} e^{-\beta S}=\sum_{\hbox{\small conf.}} \prod_p M\left(U_p \right)\;, \hbox{with } \;\; M(U_p)=\exp\{-\beta/2\left(\hbox{tr}(U_p)+\hbox{tr}(U_p^{-1}) \right)\}\;.
\end{equation}
In above equation $M(U_p)$ is the local weight for the model. Due to the invariance of the local action under cyclic permutation, the local weight is also invariant under this cyclic permutation, which makes it a class function, $M:G\rightarrow \mathbb{C}$ (equivalently, $M(g)=M(hgh^{-1})$). This construction can be generalized by choosing $M(g)$ as being any class function $M:G\rightarrow \mathbb{C}$. Moreover, we can associate local weights, $\Delta(l)$, to the edges $l$ of the lattice, such that the partition function is now given by
\begin{equation}
Z=\sum_{\hbox{\small conf.}}\prod_p M(p) \prod_l \Delta(l)\;.\label{pf}
\end{equation}
We can reproduce the usual lattice gauge theories by making appropriate choices for $M(p)$ and $\Delta(l)$ and also generate partition functions that are still gauge invariant but do not represent a physical model. The partition function in equation (\ref{pf}) is called the partition function of a generalized lattice gauge theory. Starting from this partition function we can obtain a transfer matrix whose logarithm gives us Hamiltonian operators, and thus dynamical quantum models defined over the Hilbert space $\mathcal{H}$ defined before. These quantum models are parametrized by functions of the parameters of the generalized lattice gauge theories. In~\cite{p1} it was shown that the quantum double Hamiltonians of Kitaev, of which the toric code is a special case, can be obtained from this approach. In other words it was shown how to embed such models in the parameter space of these generalized lattice gauge theories.

In~\cite{p1} it was shown that this partition function can be build out of the structure constants of an involutory Hopf algebra $\mathcal{A}$ and a 3-manifold of the form $\Sigma\times S^1$, where $\Sigma$ is some compact 2-manifold and $S^1$ is the 1-dimensional sphere. We did not consider all possible deformations of the generalized lattice gauge theory partition function in~\cite{p1}, working only with a specific kind of deformation (one parameter deformation) of the gauge theory partition function. Now we will allow other deformations by letting the parameters be any element of the center of the algebra $\mathcal{A}$ and it's dual algebra, $\mathcal{A}^*$. We only work with group algebras $\mathbb{C}G$ of a discrete group $G$ here. Nevertheless the methods presented here hold for any involutory Hopf algebra. We will go through the mathematical preliminaries beginning with the definition of the group algebra $\mathbb{C}G$.

\subsection{The Group Algebra $\mathbb{C}G$}
~

The group algebra $\mathbb{C}G$ of a discrete finite group $G$ is generated by the basis elements $\{ \phi_g: g\in G\}$ indexed by the group elements and let $\{ \Psi^g: g\in G\}$ be the dual basis such that $\Psi^g(\phi_h)=\delta(g,h)$. In this basis the multiplication and co-multiplication are defined by 

\begin{eqnarray}
\phi_a.\phi_b &:=& \phi_{ab} \;\;\;\; \Rightarrow \;\;\;\; {m_{ab}}^c=\delta(ab,c) \nonumber \\
\psi^a.\psi^b &:=& \delta(a,b)\psi^a \;\;\;\; \Rightarrow \;\;\;\; {\Delta^{ab}}_c=\delta(a,c)\delta(b,c) \nonumber 
\end{eqnarray}
and we can easily see that the unit and the co-unit of the algebra are
\begin{eqnarray}
e & = & \phi_{e} \;\;\;\; \Rightarrow \;\;\;\; e_a=\delta(a,e) \nonumber \\
\epsilon & = & \sum_{g\in G}\psi^g  \;\;\;\; \Rightarrow \;\;\;\;  \epsilon^a = 1 \nonumber 
\end{eqnarray}
where $e$ is the identity element of the group.

Finally the antipode map is defined by
\[S\left(\phi_a\right)=\phi_{a^{-1}} \;\;\;\; \Rightarrow \;\;\;\; {S_a}^b=\delta\left(ab,e\right).\]

It is not difficult to see that the group algebra structure constants satisfy all the axioms of Hopf algebras~\cite{SM, p1}. An important thing about the group algebra is the fact that it is an involutory Hopf algebra, which means that $S^2 \equiv {\bf 1}$.

As the simplest example lets consider the group algebra of $\mathbb{Z}_2$. The group $\mathbb{Z}_2$ is defined by $\mathbb{Z}_2=\{\pm 1\}$, and the product is the usual multiplication. The group algebra $\mathbb{C}\mathbb{Z}_2$ has $\{\phi_{+1}, \phi_{-1} \}$ as basis, and the product is given by
\begin{eqnarray}
\phi_{+1}\phi_{+1}=\phi_{-1}\phi_{-1}&=&\phi_{+1}\;,\nonumber \\
\phi_{-1}\phi_{+1}=\phi_{+1}\phi_{-1}&=&\phi_{-1}\;.\nonumber
\end{eqnarray}
The dual basis is $\{\Psi^{+1},\Psi^{-1}\}$, and the coproduct defined by
\begin{eqnarray}
\Psi^{+1}\Psi^{+1}&=&\Psi^{+1}\;,\nonumber\\
\Psi^{-1}\Psi^{-1}&=&\Psi^{-1}\;,\nonumber\\
\Psi^{-1}\Psi^{+1}=\Psi^{+1}\Psi^{-1}&=& 0\;.\nonumber
\end{eqnarray}
The antipode in this case is trivial $S\equiv \mathbb{1}$. 

\subsection{Constructing a Partition Function with an Involutory Hopf Algebra}
\label{sec-2.2}
~

 Consider an oriented $3D$ cubic lattice as a triangulation of a closed $3D$ manifold of the form $\Sigma\times S^1$ and let $G$ be a discrete finite group. The degrees of freedom are elements of the gauge group $G$ located on the links of this lattice. A configuration is a choice of one group element for each link of the lattice. The generalized partition function is then built by associating tensors for each face and link of the lattice which will play the role of local weights. These tensors contract with each other resulting in a scalar called generalized partition function $Z(z,\xi)$.

\begin{figure}[h!]
\centering
\subfigure[The tensor $M_{abcd}$ associated to a plaquette of the lattice.]{
\includegraphics[scale=1]{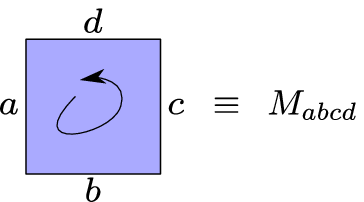} \label{tensorlattice-a}
}
\hspace{4cm}
\subfigure[The tensor $\Delta^{xyzt}$ associated to a link of the lattice.]{
\includegraphics[scale=1]{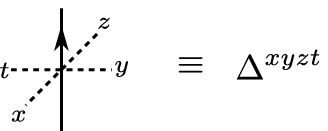} \label{tensorlattice-c}
}
\caption{Local weights associated to the plaquettes and links of the lattice.}
\label{tensorlattice}
\end{figure}

For a plaquette, whose boundary links carry the group elements $a$, $b$, $c$ and $d$, we associate a tensor $M_{abcd}$ defined by the algebra structure as shown in figure \ref{tensorlattice-a}. For a link with four plaquettes glued to it's edges, labelled by $x$, $y$, $z$ and $t$, we associate a tensor $\Delta^{xyzt}$ defined by the coalgebra structure as shown in figure \ref{tensorlattice-c}. It is very important to have an algebra and coalgebra structure so that the associated tensors can contract with each other on the closed manifold. 
These tensors are parametrized by the elements of the center of the group algebra $\mathbb{C}G$ and it's dual, $z$ and $\xi$
\begin{equation}M_{abcd}(z)=\hbox{tr}\left(z~\phi_a \phi_b \phi_c \phi_d\right),\end{equation}
and
\begin{equation}\Delta^{xyzt}(z^*)=\hbox{co-tr}\left(\xi~\Psi^x \Psi^y \Psi^z \Psi^t\right).\end{equation}
where $\hbox{tr}(\phi_g)=\vert G \vert \delta(g,e)$ is the trace in the regular representation and $\hbox{co-tr}(\Psi^g)=1,\; \forall g$. The partition function is obtained by contracting the indices of the tensors associated to the plaquettes and links. However we need to take care of the orientation of the lattice while performing this contraction. If the plaquette and link orientation matches the contraction is made directly, otherwise it is done through the antipode tensor $S_x^y$. At the end the partition function will be of the form
\begin{equation}
\label{kak}
Z(z,\xi)=\sum_{\hbox{\small indices}}\prod_p M_{abcd}(z) \prod_l \Delta^{xyzt}(\xi)\prod_{l^\prime} S_{x^\prime}^{a^\prime}\;,
\end{equation}
where the sum runs over all the contracted indices of all tensors and $l^\prime$ runs over the links with mismatching orientations.

As an example take $G=\mathbb{Z}_2$ and choose $z^g=\frac{1}{2}\left(e^{\beta} \phi_{+1} + e^{-\beta}\phi_{-1} \right)$ and $\xi^g=\epsilon$. Then the tensors will be of the form
\begin{eqnarray}
M_{abcd} &=& \hbox{tr}\left(z^g\phi_{abcd}\right)=\frac{1}{2}\hbox{tr}\left((e^{\beta}\phi_{+1}+e^{-\beta}\phi_{-1})\phi_{abcd}\right)\nonumber \\
 &=& \frac{1}{2}\left(e^{\beta}\hbox{tr}(\phi_{abcd})+e^{-\beta}\hbox{tr}(\phi_{-abcd})\right)\nonumber \\
 &=& e^{\beta(abcd)}\;;\label{bluhh}\\
\Delta^{xyzt} &=& \hbox{co-tr}\left((\xi^g \Psi^x \Psi^y \Psi^z \Psi^t\right)= \hbox{co-tr}\left(\Psi^x \Psi^y \Psi^z \Psi^t\right)\nonumber \\
&=& \delta(x,y)\delta(x,z)\delta(x,t)\;.\label{bluh} 
\end{eqnarray}
Note that $abcd$ is the holonomy $U_p$ of the plaquete $p$. In this case, due to Eq. \ref{bluh}, the sum over indices in the partition function in Eq. \ref{kak} will reduce to a sum of indices of the tensor $M_{abcd}$ (for all plaquettes), in other words, it will become a sum over configurations and the partition function will take the form
\begin{equation}
Z(z^g,\xi^g)=\sum_{\hbox{\small indices}}\prod_p M_{abcd}(z^g)\prod_l \Delta^{xyzt}(\epsilon)=\sum_{\hbox{\small conf.}}\prod_p e^{\beta (U_p)}\;,
\end{equation}
that is exactly the partition function of the lattice gauge theory defined in Eq. \ref{kaka}.

\begin{figure}[h!]
\begin{center}
		\includegraphics[scale=1]{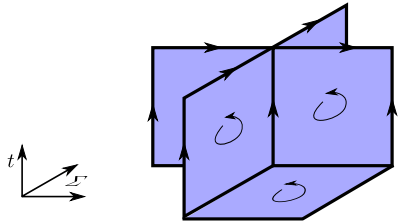}
	\caption{Time and spacelike direction.}
\label{fixingorientation}
\end{center}
\end{figure}

We can now extract a transfer matrix $U(z,\xi)$ out of this partition function, but first we have to make a distinction between timelike and spacelike directions of the 3D lattice, as shown in figure \ref{fixingorientation} for a small piece of the lattice.
Since there is now a distinction between the timelike and spacelike parts there is nothing forcing the parameters of the timelike and spacelike plaquette weights to be the same. So for a more general description we have all spacelike plaquette weights parametrized by the element $z_S$ \footnote{The parameters $z_S$ need not be the same for all the spacelike plaquettes. This shows that this construction need not obey translational invariance. This fact will be exploited to obtain disordered Hamiltonians in the next section. The same fact holds for the other parameters as well.} while the timelike ones are parametrized by $z_T$. In a similar manner the spacelike and timelike link weights are parametrized by $\xi_S$ and $\xi_T$ respectively. Therefore the partition function is now a function of the form $Z(z_S,z_T,\xi_S,\xi_T)$ \footnote{The partition function also depends on the group $G$ and the lattice $\mathcal{L}$, so it is actually a function of the form $Z(G,\mathcal{L},z_S,z_T,\xi_S,\xi_T)$. For brevity we denote it just as $Z(z_S,z_T,\xi_S,\xi_T)$.}.

\subsection{The Transfer Matrix}
~

From the partition function we have just defined we can get a transfer matrix $U$ such that its trace is equal to the partition function. This transfer matrix may depend on the same parameters as the partition function, namely $U=U(z_S,z_T,\xi_S,\xi_T)$. The operator $U$ acts on the links of the 2-dimensional lattice where the quantum states lives. A local Hilbert space $\mathcal{H}_l$ associated to each link $l$ (with basis $\{ \vert g \rangle_l: g\in G\}$). The Hilbert space of the full system is given by $\mathcal{H}=\mathcal{H}_1 \otimes\mathcal{H}_2 \otimes \cdots \otimes\mathcal{H}_N$ and a vector of this space is a linear combination of vectors of the form
\begin{equation}\vert g_1\rangle_1\otimes\vert g_2 \rangle_2\otimes \cdots\otimes \vert g_{N}\rangle_N\;.\end{equation}

\begin{figure}[h!]
\begin{center}
		\includegraphics[scale=1]{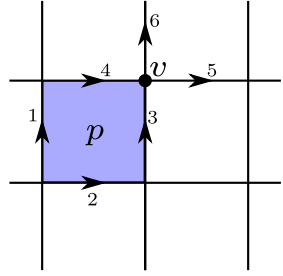}
	\caption{A plaquette and a vertex of the lattice.}
\label{linknum}
\end{center}
\end{figure}

The procedure to get such a transfer matrix was shown in~\cite{p1} using a diagrammatic notation to manage the tensors that builds the partition function and also the transfer matrix. We are now considering a more general parametrization than the one considered in \cite{p1}. The way to get the transfer matrix is similar to the procedure shown in~\cite{p1}, resulting in the same operators as in~\cite{p1} but now in a bigger parameter space. We just write down the results in what follows. The most general transfer matrix including all the parameters is given by 
\begin{equation}U(z_S,z_T,\xi_S,\xi_T)=\prod_p B_p(z_S)\prod_l C_l(z_T,\xi_S)\prod_v A_v(\xi_T)\;,\end{equation}
where $B_p(z_S)$ is an operator which acts on the links at edge of the plaquette $p$, $A_v(\xi_T)$ an operator which acts on the links sharing the vertex $v$ and $C_l(z_T,\xi_S)$ is an operator which acts on a single link. The operators $B_p(z_S)$ and $A_v(\xi_T)$ are called the plaquette and vertex operators, as before. All the parameters in $U(z_S,z_T,\xi_S,\xi_T)$ are central elements of $\mathcal{A}$ and $\mathcal{A}^*$. It can be seen in~\cite{JamesGordon} that the central elements of $\mathbb{C}G$ are written in terms of the conjugacy classes $[C]$ of $G$ as
\begin{equation}z=\sum_{C}\beta^C z_C\;,\;\;\hbox{with}\;\; z_C=\sum_{g\in[C]}\phi_g\;,\end{equation}
while the central elements of $\mathcal{A}^*$ we write in terms of the irreducible representations $R$ of $G$ as
\begin{equation}\xi=\sum_R a_R \xi^R\;,\;\; \hbox{with} \;\; \xi^R=\sum_{g\in G}\chi_R(g)\Psi^g\;,\end{equation}
where $\chi_R(g)$ is the trace of $g$ in the representation $R$. Hence the parameters of the transfer matrix can be written as
\begin{eqnarray}
z_S=\sum_C \beta^C z_C  &\hbox{and}& z_T=\sum_C b^C z_C \label{zespar} \\
\xi_S=\sum_R \alpha_R \xi^R  &\hbox{and}& \xi_T=\sum_R a_R \xi^R \label{xizespar}
\end{eqnarray}
The plaquette and vertex operators are linear functions in its parameters. That is
\begin{eqnarray}
B_p(z_S) &=& \sum_C \beta^C \underbrace{B_p(z_C)}_{B_p^C}=\sum_C \beta^C B_p^C \;, \label{bpG} \\
A_v(\xi_T) &=& \sum_R a_R \underbrace{A_v(\xi^R)}_{A_v^R}=\sum_R a_R A_v^R \;.\label{avG} 
\end{eqnarray}
where the operators $B_p^C$ and $A_v^R$ act on the links as shown in the picture in figure \ref{linknum} and are given by
\begin{eqnarray}
B_p^C &=& \vert G \vert \sum_{\{a_i \}_{i=1}^4}\delta(a_1^{-1}a_2 a_3 a_4^{-1},C) T_1^{a_1}\otimes T_2^{a_2}\otimes T_3^{a_3}\otimes T_4^{a_4}\;, \label{bpC}\\
A_v^R &=& \sum_{g\in G}\chi_R(g)R_3(g^{-1})\otimes R_4(g^{-1})\otimes L_5(g)\otimes L_6(g)\label{avR} 
\end{eqnarray}
where $\delta(g,C)=1$ if $g\in [C]$ and $\delta(g,C)=0$ if $g\notin [C]$ and the operators $L_l(\phi_g)$, $R_l(\phi_g)$ and $T_l(\Psi^g)$ are operators which act on a single link defined as
\begin{eqnarray}
L_l(\phi_g)\vert h \rangle_l &=& \vert gh\rangle_l \;, \nn \\
R_l(\phi_g)\vert h \rangle_l &=& \vert hg \rangle_l \;, \nn \\ 
T_l(\Psi_g)\vert h \rangle_l &=& \delta(g,h) \vert h\rangle_l. \; \label{lrtopss}
\end{eqnarray}
These operators are linear on its parameters, in other words, $L(z)=\sum_g z^gL(\phi_g)$ (this property also holds for $R(z)$ and $T(z^*)$. Sometimes we use the short notation $L^g=L(\phi_g)$, $R^g=R(\phi_g)$ and $T^g=T(\Psi^g)$. The link operator $C_l(z_T,\xi_S)$ can also be written in terms of the $L_l$ and $T_l$ operator as we shall see next.

The main difference between this model in the one we have considered in~\cite{p1} is the link operator $C_l(z_T,\xi_S)$. Unlike in~\cite{p1} this operator is no longer proportional to identity, now it takes the form
\begin{equation}C_l(z_T,\xi_S)=\vert G \vert T_l(\xi_S)L_l(z_T)=\vert G\vert\left(\sum_R \alpha_R \underbrace{T_l(\xi^R)}_{T_l^R} \right)\left(\sum_C b^C \underbrace{L_l(z_C)}_{L_l^C} \right)\;,\end{equation}
where
\begin{eqnarray}
L_l^C=R_l^C &=&\sum_{g\in [G]}L_l^g=\sum_{g\in [G]}R_l^g \label{LC} \\
T_l^R &=& \sum_{g\in G}\chi_R(g)T_l^g \label{TR}
\end{eqnarray}
Thus the final expression for the transfer matrix is
\begin{equation}
U(z_S,z_T,\xi_S,\xi_T)=\vert G \vert^{n_l}\prod_p\left(\sum_C \beta^C B_p(z_C)\right)\prod_l \left[ \left(\sum_R \alpha_R T_l^R\right)\left(\sum_C b^C L_l^C \right)\right]\prod_v \left( \sum_R a_R A_V^R\right)\;,
\label{finalU}
\end{equation}
where the parameters on the left hand side are related to the coefficients on the right hand side by the equations \eqref{zespar} and \eqref{xizespar}.

\subsection{Algebra of the operators}
~

In order to find the algebra that the operators $B_p$, $A_v$ and $C_l$ satisfy one should first look at the algebra of the operators $L_l$, $R_l$ and $T_l$.
It is not difficult to see that the following relations holds
\begin{equation}
\begin{array}{lcr}
L_l^g L_l^h = L_l^{gh}  &  R_l^g R_l^h = R_l^{hg}  &  T_l^g T_l^h = \delta(g,h)T_l^g \\
L_l^gR_l^h = R_l^h L_l^g  &  L_l^gT_l^h=T_l^{hg}L_l^g  &  R_l^g T_l^h = T_l^{hg}R_l^g 
\end{array}
\label{LRTrelation}
\end{equation}
Now using Eq.(\ref{LRTrelation}) and the orthogonality relations on the characters $\chi_R(g)$~\cite{JamesGordon} we can show the algebra of the operators which build the transfer matrix, namely $B_p^C$, $A_v^R$, $L_l^R$ and $T_l^R$.

The set of operators $\{B_p^C\}$ is a complete basis of orthogonal projectors which generate the plaquette operators, which means
\begin{equation}
B_p^C B_p^{C^\prime} =\vert G \vert \delta(C,C^\prime)B_p^C\;, \;\; \hbox{and}\;\; \vert G \vert^{-1} \sum_C B_p^C = {\bf 1}\;.
\label{BpCort} 
\end{equation}
Same way the set of $\{A_v^R\}$ is a complete basis of orthogonal projectors which generate the vertex operator, in other words
\begin{equation}
A_v^R A_v^{R^\prime} =\delta(R,R^\prime)A_v^R\;, \;\; \hbox{and}\;\; \sum_R A_v^R = {\bf 1}\;.
\label{AvCort} 
\end{equation}
The plaquette and vertex operator still commute for any choice of the parameters $z_S$ and $\xi_T$, so we can write
\begin{equation}\left[B_p^C,A_v^R\right]=0\;\;\; \Rightarrow \;\;\; \left[B_p(z_S),A_v(\xi_T)\right]=0\;,\; \forall z_S, \xi_T.\end{equation}
The set of operators $\{L_l^C\}$ and $\{T_l^R\}$ are also complete sets of  orthogonal projectors ($L_l^C L_l^{C^\prime}=\delta(C,C^{\prime})L_l^C$ and $T_l^R T_l^{R^\prime}=\delta(R,R^{\prime})T_l^R$), however the operator $T_l^R$ does not commute with the link operator but it does commute with the plaquette operator, in the same way as the operator $L_l^C$ does not commute with the plaquette operator but it does commute with the vertex operator. Thus we can write 
\begin{equation}
[T_l^R,B_p^C]=0 \;, \;\;\; [T_l^R,A_v^R]\neq 0 \;, \;\;\; [L_l^C,B_p^C]\neq 0\;, \;\;\; [L_l^C,A_v^R]=0 \;.
\end{equation}
Therefore, the link operator $C_l(z_T,\xi_S)$ commutes with the plaquette and vertex operators only for some choices of the parameters $z_T$ and $\xi_S$, which means the quantum model obtained from the transfer matrix containing this link operator is in general not solvable.

\section{Examples of Models from the Transfer Matrix}

The fully parametrized transfer matrix obtained in the previous section helps us construct a number of other interesting models. We have seen that the quantum double Hamiltonians are only one special class of models in this parameter space. Here we will look at what other possibilities exist in the extended parameter space. The first set of examples consist of disordered quantum double Hamiltonians. These are quantum double Hamiltonians which do not have translational invariance. This is due to the appearance of coefficients, for the vertex and plaquette terms, which are not constant but depend on the vertex $v$ and plaquette $p$. These models continue to be in the quantum double phase. The trace of the transfer matrix for these Hamiltonians also help us obtain their partition functions. 

We can produce solvable models even in the presence of the parameters $z_T$ and $\xi_S$. We discussed a class of such models in \cite{p3} where we exhibited exactly solvable models which continued to remain in the quantum double phase described by modified vertex operators or plaquette operators. Here we show another class of models that can be obtained from the transfer matrices of lattice gauge theories that continue to remain exactly solvable but the excitations are ``partially'' confined with respect to those of the original quantum double model. By this we mean that the deconfined quantum double excitations are now confined up to a few steps due to the addition of an extra term in the Hamiltonian. The number of steps for which they are confined can be controlled by adding the appropriate term in the Hamiltonian. We will call these models $n$-step confined models. These models comprise our second set of examples. They are examples where the self-duality of the quantum double Hamiltonians is broken. We also discuss the ground states, it's degeneracy apart from the excited states of the model.

Finally we write down models that are obtained by using the remaining two parameters $z_T$ and $\xi_S$ along with $z_S$ and $\xi_T$. The transfer matrix is now significantly modified as new single qudit operators, acting on individual links, appear. They do not commute with the vertex and plaquette operators in general. However for certain special values of parameters they commute with products of vertex and plaquette operators as we shall see when we consider these models later in this section. The models obtained at these values are quantum double Hamiltonians perturbed by generalized ``magnetic'' fields. They have the interpretation of magnetic fields in the case the input algebra $\mathcal{A}$ is $\mathbb{C}(Z_2)$. For the remaining values of the parameters we obtain more complicated terms in the Hamiltonian which we will briefly touch upon. These models are not exactly solvable and are outside the phase described by the quantum double Hamiltonians, at least for sufficiently large values of the parameters. 

\subsection{Disordered Quantum Double Hamiltonians (QDH)} 

The transfer matrices used to obtain these models only use $z_S$ and $\xi_T$. The other two parameters are set to $z_T=\eta$ and $\xi_S=\epsilon$ for all the timelike plaquettes and spacelike links. To obtain the disordered QDH models we associate a different central element of the algebra and it's dual to every spacelike plaquette and timelike link respectively. This leads to the following transfer matrix
\beq U\left(\mathcal{A}, z_{S,p}, \xi_{T,v}\right) = \prod_p B_p(z_{S,p})~\prod_v A_v(\xi_{T,v}) \eeq
where $z_{S,p}$ and $\xi_{T,v}$ are the parameters for the plaquette $p$ and vertex $v$ respectively. The plaquette and vertex operators commute with each other in this transfer matrix and each of the operators is a sum of projectors. Thus it is easy to take the logarithm of these matrices to obtain the disordered QDHs. 

Let us look at these Hamiltonians in the case when $\mathcal{A}=\mathbb{C}(Z_2)$. Denote the basis elements of $\mathbb{C}(Z_2)$ by $\{\phi_1, \phi_{-1}\}$ with $\phi_{-1}^2=\phi_1$, and the basis elements of the dual $\mathbb{C}(Z_2)^*$ by $\{\psi^1 , \psi^{-1}\}$ with the product $\psi^i.\psi^j =\delta(i,j)\psi^i$. The central elements of $\mathbb{C}(Z_2)$ and $\mathbb{C}(Z_2)^*$ can be written as $z= \alpha_1\phi_1 + \alpha_{-1}\phi_{-1}$ and $\xi=\beta_1\psi^1 + \beta_{-1}\psi^{-1}$ respectively. By assigning each spacelike plaquette and timelike link with $z_{S,p}$ and $\xi_{T,v}$ respectively we obtain the transfer matrix as
\beq U\left(\mathbb{C}(Z_2), z_{S,p}, \xi_{T,v}\right) = \prod_p B_p(z_{ S,p})~\prod_vA_v(\xi_{T,v}) \eeq
where the plaquette operators are given by
\beq B_p(z_{S,p}) = \alpha_{1,p} B_p^1 + \alpha_{-1,p} B_p^{-1} \eeq
with 
\beq B_p^{\pm 1} = \frac{1 \pm \sigma^z_{i_1}\otimes\sigma^z_{i_2}\otimes\sigma^z_{i_3}\otimes\sigma^z_{i_4}}{2} \eeq
and the vertex operators are given by 
\beq A_v(\xi_{T,v}) = \left(\frac{\beta_{1,v}+\beta_{-1,v}}{2}\right) A_v^1 + \left(\frac{\beta_{1,v}-\beta_{-1,v}}{2}\right) A_v^{-1} \eeq
with 
\beq A_v^{\pm 1} = \frac{1 \pm \sigma^x_{j_1}\otimes\sigma^x_{j_2}\otimes\sigma^x_{j_3}\otimes\sigma^x_{j_4}}{2}. \eeq
The action is shown in figure (\ref{linknum}). 

The disordered QDH can now be written as 
\beq \label{disH}H = \sum_v \left(\ln\left(\frac{\beta_{1,v}+\beta_{-1,v}}{2}\right) A_v^1 + \ln\left(\frac{\beta_{1,v}-\beta_{-1,v}}{2}\right) A_v^{-1}\right) + \sum_p \left(\ln\alpha_{1,p} B_p^1 + \ln\alpha_{-1,p} B_p^{-1}\right).\eeq
As it can be seen this Hamiltonian breaks translational invariance and is made up of a sum of commuting projectors. When translational invariance is restored we recover the familiar toric code Hamiltonian. 

\begin{figure}[h!]
\centering
\subfigure[Non-contractible loops on the direct and on the dual lattice.]{
\includegraphics[scale=1]{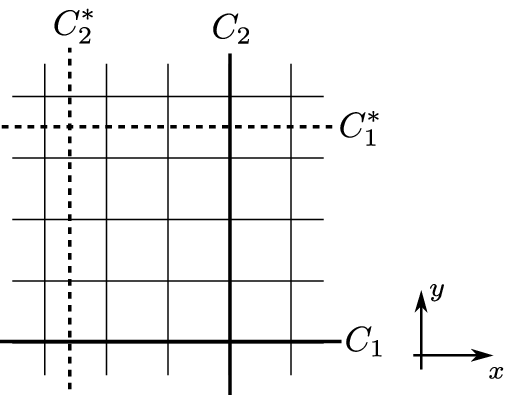} \label{noncontract-a}
}
\hspace{2cm}
\subfigure[Non-contractible loops on a torus.]{
\includegraphics[scale=1]{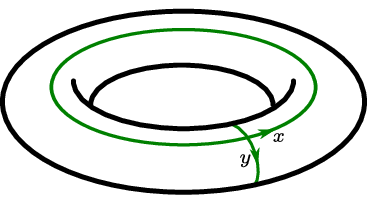} \label{noncontract-b}
}
\caption{Non-contractible loops.}
\label{noncontract}
\end{figure}

The ground state for this Hamiltonian is easily obtained by projecting on to the smallest of $\beta_{1,v}$ and $\beta_{-1,v}$ for every vertex $v$ and $\alpha_{1,p}$ and $\alpha_{-1,p}$ for every plaquette $p$. The ground state degeneracy is the same as the usual toric code and the winding operators, 
\beq\label{xw} X_{C_1^*, C_2^*, (C_1^*, C_2^*)} = \prod_{j\in C_1^*, C_2^*, (C_1^*, C_2^*)} \sigma^x_j \eeq
\beq \label{zw}Z_{C_1, C_2, (C_1, C_2)} = \prod_{k\in C_1, C_2, (C_1, C_2)}\sigma^z_k \eeq
where the non-contractible loops $C_1, C_2, C_1^*, C_2^*$ are defined on the direct and dual lattice respectively as shown in the figure \ref{noncontract},
commute with the Hamiltonian.

\begin{figure}[h!]
\begin{center}
		\includegraphics[scale=1]{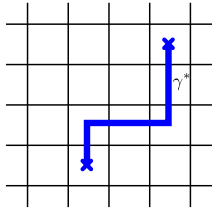}
	\caption{Deconfined flux excitations.}
\label{deconfstepflux}
\end{center}
\end{figure}

The excitations correspond to the other value of the coefficients for each vertex and plaquette. The string operators (ribbon operators in the case of the non-Abelian groups~\cite{Del2}) creating the excitations are the same as in the QDH case. For the specific case of $\mathbb{C}(Z_2)$ we have the string operators, creating charge or vertex excitations at the end points of the string $\gamma$ along the direct lattice, as 
\beq \label{zs}V_\gamma = \prod_{j\in \gamma} \sigma^z_j \eeq 
and those of the fluxes or plaquette excitations at the end points of the string $\gamma^*$ along the dual lattice as
\beq\label{xs} P_{\gamma^*} = \prod_{k\in \gamma^*} \sigma^x_k. \eeq
These are shown in figure \ref{deconfstepflux}. As is well known these excitations are deconfined by which we mean that there is no cost in energy for moving them around by stretching the string creating them. Moreover the fusion rules and braiding statistics are the same as in the translationally invariant toric code. Thus we conclude that the disordered QDH given in Eq. (\ref{disH}) continues to remain in the toric code phase.

\subsection{ Quantum Double Hamiltonian with $n$-Step Confined Excitations}

We first consider the case with $n=2$. To obtain this we write down an example of exactly solvable Hamiltonian made up of the QDH vertex and plaquette operators along with new terms made of these operators which have 2-step confined low energy excitations unlike the QDH case where all the excitations are completely deconfined. The transfer matrix used to obtain these models contain only the $z_S$ and $\xi_T$ parameters.

The transfer matrix can be written as
\beq U\left(\mathcal{A}, z_S, \xi_T\right) = \prod_pB_p(z_S)~\prod_vA_v(\xi_T). \eeq
In the case of $\mathbb{C}(Z_2)$ we can write down the Hamiltonian with the 2-step confined charges and fluxes as follows
\bea H & = & \nn \sum_v \left(\alpha_1 A_v^1 + \alpha_{-1}A_v^{-1}\right) + \sum_p\left(\beta_1 B_p^1 + \beta_{-1} B_p^{-1}\right)  \\  & + & \nn  \sum_{<ij>} \left(\alpha_1' A_{v_i}^1A_{v_j}^1 + \alpha_2' A_{v_i}^1A_{v_j}^{-1} + \alpha_3' A_{v_i}^{-1}A_{v_j}^1 +\alpha_4'A_{v_i}^{-1}A_{v_j}^{-1}\right) \\ &  + &  \sum_{<i^*j^*>} \left(\beta_1' B_{p_i^*}^1B_{p_j^*}^1 + \beta_2' B_{p_i^*}^1B_{p_j^*}^{-1} + \beta_3' B_{p_i^*}^{-1}B_{p_j^*}^1 +\beta_4'B_{p_i^*}^{-1}B_{p_j^*}^{-1}\right) \eea
where $<ij>$ and $<i^*j^*>$ are nearest neighbor vertices in the direct and dual lattices respectively. All the terms in this Hamiltonian commute with each and other and are sums of projectors. The ground states are given by the usual toric code Hamiltonian. The degeneracy does not change as the winding operators in the toric code case, given by Eq.(\ref{xw}) and Eq.(\ref{zw}), continue to commute with this Hamiltonian and thus help create the new states from a given ground state. In particular on a torus the degeneracy is four. 

\begin{figure}[h!]
\centering
\subfigure[The pair of charges with energy $E$.]{
\includegraphics[scale=1]{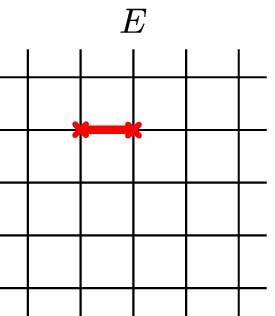} \label{confstep-a}
}
\hspace{1cm}
\subfigure[Moving by one step increases the energy by $\Delta E$.]{
\includegraphics[scale=1]{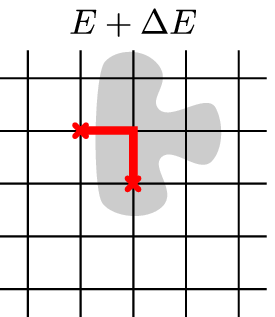} \label{confstep-b}
}
\hspace{1cm}
\subfigure[Moving one more step outside the shaded region does not cost any energy. ]{
\includegraphics[scale=1]{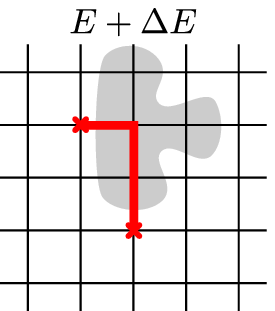} \label{confstep-c}
}
\hspace{1cm}
\subfigure[Once outside the shaded region the charges are deconfined.]{
\includegraphics[scale=1]{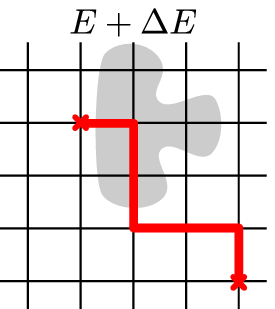} \label{confstep-d}
}
\caption{The 2-step confined charge excitations.}
\label{confstep}
\end{figure}

\begin{figure}[h!]
\centering
\subfigure[The pair of fluxes with energy $E$.]{
\includegraphics[scale=1]{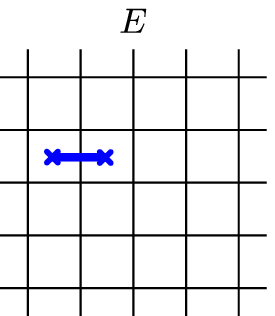} \label{confstepflux-a}
}
\hspace{1.5cm}
\subfigure[Moving the flux by one step increases the energy by $\Delta E$.]{
\includegraphics[scale=1]{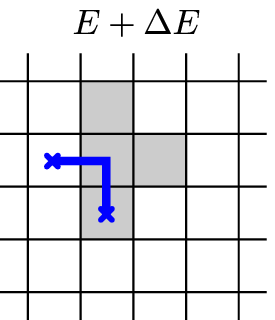} \label{confstepflux-b}
}
\hspace{1.5cm}
\subfigure[Once outside the shaded region they are deconfined.]{
\includegraphics[scale=1]{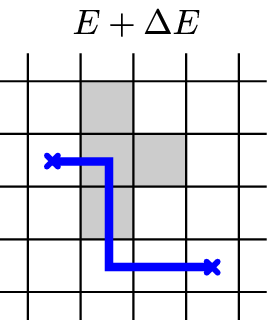} \label{confstepflux-c}
}
\caption{The 2-step confined flux excitations.}
\label{confstepflux}
\end{figure}

The interesting feature of this model occurs when we look at the excitations. As in the toric code case the string operators creating charge and flux excitations are given by Eq.(\ref{zs}) and Eq.(\ref{xs}) respectively. However in this model when we create a charge or flux excitation we also excite the direct or the dual link given by the $A_{v_i}^1A_{v_j}^1$ or the $B_{p_i^*}^1B_{p_j^*}^1$ \footnote{ We assume that the coefficients of these two terms in the Hamiltonian are the smallest and hence violating these will lead to excitations. There is no loss of generality in making this assumption.} term respectively. This creates link excitations along the string where the operator given by Eq.(\ref{zs}) or Eq.(\ref{xs}) acts. However the creation of these link excitations occurs only for two steps after which they are deconfined as in the toric code case. Thus we say that these excitations as 2-step confined excitations. These partially confined excitations are shown in figure (\ref{confstep}) and (\ref{confstepflux}).  
 
It is easy to see that the dyonic excitations are also confined in a similar manner. Thus we have a model based on lattice gauge theory which is exactly solvable, has ground state degeneracy and has excitations which are confined up to two steps or can be thought of as being partially confined.
 
 There is a natural way to increase the number of steps for which these particles are confined. This is achieved by coupling more number of vertex and plaquette operators. For example to obtain three-step confinement we add terms of the form $A_{v_i}^1A_{v_j}^1A_{v_k}^1$ where $i, j, k$ are nearest neighbors on the direct lattice. The corresponding plaquette terms are $B_{p_i^*}^1B_{p_j^*}^1B_{p_k^*}^1$ where $i^*, j^*, k^*$ are nearest neighbor vertices on the dual lattice. The energy of these excitations are more when compared to the original QDH as we violate more terms in the Hamiltonian to obtain them. This argument can easily be extended to any number of steps. The corresponding figures of the $n$-step confined excitations will have a larger shaded region where they are confined when compared to the $n=2$ case. 

This model can be constructed on any triangulation of a two dimensional lattice keeping the confinement properties unchanged. This is easy to see as the additional terms of pairing neighboring vertex and plaquette operators is independent of triangulation. The model can also be easily extended to all group algebras and more generally to all involutory Hopf algebras without any obstacle.   

In the case of $\mathbb{C}(Z_n)$ we can also add the parameters $z_T$ and $\xi_S$, to include the single qudit terms in the transfer matrix and keep the above properties of confinement. We will illustrate why this is so for $\mathbb{C}(Z_2)$. The argument extends easily for other $n$. We have the following relation 
\beq A_{v_1}^1A_{v_2}^1 T_l A_{v_1}^1A_{v_2}^1 = A_{v_1}^1A_{v_2}^1 \eeq
where $T_l = \frac{1+\sigma^z_l}{2}$ and $v_1$, $v_2$ are the end points of the link $l$. This reduces the model to the previous Hamiltonian showing confinement. From this identity we see that we can include the parameters $\xi_S$ in the transfer matrix without any effect to the properties of the Hamiltonian considered earlier.  

Also note that the braiding and fusion rules do not change in the $n=2$ case. The only difference from the usual toric code is that the energy of the quasiparticle excitations is now higher and there is an energy cost to move them up to one step in the lattice. Since the topological data is the same we conclude that this model is in the same phase as the toric code. By increasing $n$ we could enter a different phase, namely a confined phase, in the thermodynamic limit. 

\subsection{Perturbed QDHs}

These are constructed out of transfer matrices which include the parameters $z_T$ or $\xi_S$. Their inclusion introduces single qudit operators $L_l(z_T)$ and $T_l(\xi_S)$ on the link $l$ respectively. These terms do not commute in general with the vertex and plaquette operators thereby making the process of taking their logarithms and hence obtaining the Hamiltonian difficult. However for certain parameters we can still take the logarithm to obtain a Hamiltonian. We will illustrate this in the case of $\mathcal{A}=\mathbb{C}(Z_2)$. 

Let us include the parameter $z_T$ into the transfer matrix. This brings in the operator 
\beq L_l(z_T) = \left(\frac{x_1+x_{-1}}{2}\right) \left(1+\sigma^x_l\right) + \left(\frac{x_1-x_{-1}}{2}\right) \left(1-\sigma^x_l\right) \eeq
when $z_T = x_1\phi_1 + x_{-1}\phi_{-1}$. When $L_l(z_T) = \sigma^x_l$ we can have a product of two plaquette operators adjacent to the link $l$ commute with $L_l(z_T)$. We can then take the logarithms to obtain the following Hamiltonian
\beq H = \sum_v\left(ln(\alpha_1) A_v^1 + ln(\alpha_{-1}) A_v^{-1}\right) + \sum_p\left(ln(\beta_1) B_p^1 + ln(\beta_{-1}) B_p^{-1}\right) + i\pi\left(\frac{1-\sigma^x_l}{2}\right). \eeq
This model resembles adding magnetic perturbations to the QDH for the simplest case of $\mathbb{C}(Z_2)$. The dual version of this involves using the parameter $\xi_S$ in the transfer matrix to obtain the magnetic field operator $\sigma_l^z$ on the links. 

 \section{Outlook}

A systematic procedure was used to obtain the models with interesting properties which included studying all possible ways of taking the logarithms of the transfer matrices of generalized lattice gauge theories. In~\cite{p1} this led to quasi-topological phases with increased ground state degeneracy occurring due to condensed excitations of the QDH. In this paper another way of taking the logarithm led to partially confined excitations. We went further to include other parameters in the transfer matrix which took us away from the topologically ordered phases. 

The state sum procedure used in~\cite{p1} and further explored here can be thought of as a method to construct the quantum double of a given input algebra. Using this principle we can use other inputs to obtain the quantum doubles of more general objects leading to more interesting models. One such input are groupoid algebras which are examples of quantum groupoids. Such considerations lead to confined excitations in pure lattice gauge theories~\cite{p4}. In~\cite{LChang} the quantum doubles of weak Hopf algebras were constructed. These reproduced the Levin-Wen models. Our considerations will embed the Levin-Wen models in the parameter space of lattice gauge theories based on these weak Hopf algebras. The weak Hopf algebras used in~\cite{LChang} were the ones constructed from a unitary fusion category by Kitaev and Kong in~\cite{Kit2}. In particular by using weak Hopf algebras of~\cite{Kit2} as inputs in the construction of~\cite{p1} we will be able to obtain the operators creating excitations in the Levin-Wen models and finally we can confine and condense these quasi particles by using the methods in this paper. 

The state sum construction can be used to construct the transfer matrices of lattice theories with gauge and matter fields by adding matter degrees of freedom, acted upon by the gauge fields, on the vertices of the triangulated lattice. This construction was shown to produce exactly solvable quantum models in one and two dimensions in~\cite{p5}. The methods of this paper when applied to the transfer matrix constructed in~\cite{p5} is bound to give many new interesting models.

\section{Acknowledgements}
The authors would like to thank FAPESP for support of this work.

\end{document}